\documentclass[11pt]{article}

\usepackage{amsmath}
\usepackage{graphicx}
\usepackage{indentfirst}
\usepackage{amssymb}
\usepackage{cite}
\usepackage{color}
\usepackage{subfigure}

\begin{document}

\title{\bf{Thermodynamics in Rotating Anti-de Sitter Black Holes with Massive Scalar Field in Three Dimensions}}

\date{}
\maketitle

\begin{center}
\author{Bogeun Gwak}$^a$\footnote{rasenis@dgu.ac.kr}\\

\vskip 0.25in
$^{a}$\it{Division of Physics and Semiconductor Science, Dongguk University, Seoul 04620,\\Republic of Korea}\\
\end{center}
\vskip 0.6in

{\abstract
{We investigate the tendency in variations of the dual CFT$_2$ when a rotating AdS$_3$ black hole changes by fluxes transferred by the scattering of a massive scalar field according to the AdS/CFT correspondence. The conserved quantities of the black hole are definitely constrained by the extremal condition. Moreover, the laws of thermodynamics provide a direction for the changes in the conserved quantities, so the black hole cannot be extremal under the scattering. This is naturally preferred. By the relationship between the rotating AdS$_3$ black hole and dual CFT$_2$, we find that such changes in the black hole constrain variations in eigenstates of the dual CFT$_2$. Furthermore, the tendency in the variations is closely related to the laws of thermodynamics.}}

\thispagestyle{empty}
\newpage
\setcounter{page}{1}

\section{Introduction}\label{sec1}

Black holes are a type of compact objects with an event horizon, from which all particles, including light, cannot escape. Owing to the effect of the event horizon, an observer located outside the black hole is unable to detect emissions from the horizon. However, considering the quantum effects, there is a small emission from black holes; this emission is  called Hawking radiation\cite{Hawking:1974sw,Hawking:1976de}. Therefore, black holes can be treated as radiative objects possessing Hawking temperature. Furthermore, black holes are known to have an irreducible mass that does not decrease during any physical process\cite{Christodoulou:1970wf,Bardeen:1970zz,Christodoulou:1972kt,Smarr:1972kt}. Focusing on this irreducible nature, which is similar to the second law of thermodynamics, the entropy of a black hole was determined to be proportional to the surface area of the horizon\cite{Bekenstein:1973ur,Bekenstein:1974ax}. Hence, a black hole can be considered as a system having Hawking temperature and Bekenstein--Hawking entropy. We also construct the laws of thermodynamics applied to black holes, in terms of such thermodynamic variables.

Thermodynamic properties of black holes are defined at the horizon. In this case, the horizon acts as an effective surface of the black hole, because of the nature of the horizon, which divides the black hole into its inside and outside. In particular, the horizon hides the curvature singularity from the outside observer. As a naked singularity observed without horizons can cause the breakdown of a causal structure, the singularity is conjectured to be hidden by the horizon. This is known as the weak cosmic censorship conjecture (WCCC)\cite{Penrose:1964wq,Penrose:1969pc}. The first test to the WCCC is Wald's gedanken experiment for the Kerr black hole by adding a particle\cite{Wald:1974ge}. Thereafter, the WCCC has been actively studied for various black holes and tested channels. Based on adding a particle, the WCCC is also investigated considering several effects such as self-force and the back-reaction effects\cite{Hubeny:1998ga,Jacobson:2009kt,Barausse:2010ka,Isoyama:2011ea,Colleoni:2015afa,Gwak:2015fsa,Sorce:2017dst,Gao:2012ca,Rocha:2014jma,Cardoso:2015xtj,Siahaan:2015ljs,Gwak:2016gwj,Revelar:2017sem,Gim:2018axz,Yu:2018eqq,Zeng:2019jrh,Wang:2019jzz,He:2019kws,Hu:2019zxr}. Furthermore, the WCCC can be studied based on the scattering of an external field, because the back-reacted black hole alters its states depending on the modes of the external field\cite{Hod:2008zza,Semiz:2005gs,Toth:2011ab,Natario:2016bay,Duztas:2017lxk,Gwak:2018akg,Gwak:2019asi,Duztas:2019ick,Chen:2019nsr,Natario:2019iex,Jiang:2019vww,Wang:2019bml,Gwak:2019rcz,Hong:2020zcf}. Although the external field affects the black hole, the changes in the black hole are in agreement with the laws of thermodynamics, and these laws ensure the validity of the WCCC\cite{Gwak:2015fsa,Gwak:2018akg}. Consequently, the laws of thermodynamics can be closely associated with the WCCC.

The thermodynamics of black holes is also important in terms of the anti-de Sitter (AdS)/conformal field theory (CFT) correspondence. Here, the gravity theory described in the AdS spacetime is related to the CFT in its one dimensional lower boundary\cite{Maldacena:1997re,Gubser:1998bc,Witten:1998qj,Aharony:1999ti}. According to the AdS/CFT correspondence, the properties of AdS black holes provide meaningful insights into the physics of the dual CFTs. The thermodynamics of black holes is related to that of the dual CFT. Moreover, as the Hawking radiation provides the temperature of a black hole, the dual CFT associated with an AdS black hole has a finite temperature provided by the Hawking temperature, identical to the AdS black hole in the bulk\cite{Witten:1998zw}. Furthermore, in three-dimensional AdS black holes, the generators of diffeomorphisms can be obtained such that they preserve the asymptotic boundary conditions and obey the Virasoro algebra with a specific central charge\cite{Brown:1986nw}. Thus, we can determine the dual CFT to the AdS$_3$. When the AdS$_3$ black holes are assumed to be an excitation from the zero-mass vacuum, the Cardy formula corresponds to the Bekenstein--Hawking entropy, remarkably\cite{Strominger:1997eq}. The AdS/CFT correspondence has been extended to various topics such as quantum chromodynamics (QCD)\cite{Babington:2003vm,Kruczenski:2003uq,Sakai:2005yt,Erlich:2005qh} and condensed matter theory (CMT)\cite{Hartnoll:2008vx,Hartnoll:2008kx}. In particular, AdS$_3$ black holes, which are considered in this study, are dual to holographic superconductors\cite{Jensen:2010em,Andrade:2011sx,Chang:2014jna} in terms of the AdS/CMT correspondence.

In this study, we investigate the tendency of variations of the dual CFT$_2$, while the rotating AdS$_3$ black hole changes owing to the flux transfer caused by the scattering of a massive scalar field in the gravity side. In the gravity side, variables of the black hole, such as mass and angular momentum, are bounded by the extremal condition. In particular, the laws of thermodynamics ensure that the extremal condition functions as a physical boundary. In spite of the AdS/CFT correspondence, the dual description of the extremal condition still remains unclear. As the rotating AdS$_3$ black hole is a well-studied case that is directly related to the dual CFT$_2$, we determine the implications of the dual CFT$_2$ with respect to the extremal condition, while ensuring that the second and third laws of thermodynamics for the black hole are satisfied. Furthermore, according to the third law, the temperature of the black hole cannot be zero owing to any physical processes, such as the scattering of a scalar field considered in our study. We also investigate the dual description regarding the zero-temperature bound. Therefore, the bound found in the dual CFT$_2$ will be enforced by the second law. Finally, we can obtain the relationship of the bounded behaviors between the rotating AdS$_3$ black hole and dual CFT$_2$, for the variations originating from the scattering of the massive scalar field.

The remainder of this paper is organized as follows. In section\,\ref{sec2}, we briefly review the relationship between the rotating AdS$_3$ black hole and its dual CFT$_2$. In section\,\ref{sec3}, the solution to the scattering of a massive scalar field at the horizon of the black hole is described. In section\,\ref{sec4}, the effects of the scalar field on the black hole are described in terms of the fluxes of the scalar field. Section\,\ref{sec5} describes the variations in the dual CFT$_2$ originating from the changes in the black hole caused by the scalar field. The results are briefly summarized in section\,\ref{sec6}.

\section{Review on Rotating AdS$_3$ Black Hole and Dual CFT$_2$}\label{sec2}

We consider the relationship between the rotating AdS$_3$ black hole and its dual CFT$_2$, under the changes caused by the scattering of a scalar field. The AdS/CFT correspondence is well constructed for rotating AdS$_3$ black holes. Moreover, and the entropy of the black hole according to microscopic derivation has been well explained in \cite{Brown:1986nw,Strominger:1997eq}. Here, the derivation is briefly introduced to elucidate the dual description for the AdS$_3$ black hole. The three-dimensional action with a cosmological constant is
\begin{align}\label{eq:action1}
S=\frac{1}{16\pi G} \int d^3x\sqrt{-g}(R+\frac{2}{\ell^2}),
\end{align} 
where $R$ is the curvature, and $\ell$ is the AdS radius. Considering the dual CFT$_2$, the AdS radius is assumed in the semiclassical limit such as $\ell\gg G$. The action in Eq.\,(\ref{eq:action1}) includes the AdS$_3$ spacetime in an $SL(2,R)_L\otimes SL(2,R)_R$ isometry group. To define the dual CFT$_2$ at the boundary of the AdS$_3$ bulk, the asymptotic boundary condition is given as\cite{Strominger:1997eq}
\begin{align}\label{eq:boundary1}
g_{tt}&=-\frac{r^2}{\ell^2}+\mathcal{O}(1),\quad g_{t\phi}=\mathcal{O}(1),\quad g_{tr}=\mathcal{O}(r^{-3}),\\
g_{rr}&=\frac{\ell^2}{r^2}+\mathcal{O}(r^{-4}),\quad g_{r\phi}=\mathcal{O}(r^{-3}),\quad g_{\phi\phi}=r^2+\mathcal{O}(1).\nonumber
\end{align}
Diffeomorphisms can be constructed to preserve Eq.\,(\ref{eq:boundary1}). The generators of diffeomorphisms are defined as $L_n$ and $\bar{L}_n$ with $-\infty<n<+\infty$, and these also obey the Virasoro algebra with the central charge $c=\frac{3\ell}{2G}$. This implies that the CFT$_2$ with $c=\frac{3\ell}{2G}$ is dual to the gravity theory on the AdS$_3$ bulk.

The rotating AdS$_3$ black hole with mass $M$ and angular momentum $J$ has an AdS$_3$ boundary that satisfies Eq.\,(\ref{eq:boundary1})\cite{Banados:1992wn,Banados:1992gq}. When the horizon is located at $r=r_\text{h}$, its metric is expressed  as\cite{Strominger:1997eq}
\begin{align}\label{eq:BHmetric1}
ds^2=-N^2 dt^2+\rho^2(N^\phi dt+d\phi)^2+\frac{r^2}{N^2 \rho^2}dr^2,
\end{align}
and
\begin{align}
N^2=\frac{r^2(r^2-r_\text{h}^2)}{\ell^2 \rho^2},\quad N^\phi=-\frac{4GJ}{\rho^2},\quad\rho^2=r^2+4GM\ell^2-\frac{1}{2}r^2_\text{h},\quad r_\text{h}^2=8G\ell\sqrt{M^2\ell^2-J^2}.
\end{align}
This metric is significantly different from that of the BTZ black hole in \cite{Banados:1992wn,Banados:1992gq}, because the metric in Eq.\,(\ref{eq:BHmetric1}) is transformed to associate the CFT$_2$. The rotating velocity at the horizon is
\begin{eqnarray}
\Omega_\text{h}=\frac{4GJ}{\rho_\text{h}^2},\quad \rho_\text{h}^2=\frac{1}{2}r_\text{h}^2+4GM\ell^2.
\end{eqnarray}
The Hawking temperature and Bekenstein--Hawking entropy are expressed as
\begin{eqnarray}
T_\text{H}=\frac{4G\sqrt{M^2\ell^2-J^2}}{\pi \ell \rho_\text{h}},\quad S_\text{BH}=\frac{\pi \sqrt{16GM\ell^2+2r_\text{h}^2}}{4G}.
\end{eqnarray}
As the boundary condition in Eq.\,(\ref{eq:BHmetric1}) satisfies Eq.\,(\ref{eq:boundary1}), the dual theory is still the CFT$_2$ with $c=\frac{3\ell}{2G}$. However, the metric of the AdS$_3$ spacetime is not at $M=J=0$ but at $M=-\frac{1}{8G}$. Hence, the black hole and the AdS$_3$ spacetime are locally equivalent\cite{Strominger:1997eq}. Because of the naked singularity in $-\frac{1}{8G}<M<0$, the black hole in Eq.\,(\ref{eq:BHmetric1}) is regarded as the finite mass excitation from the black hole with $M=0$. Thus, the mass and angular momentum are related to the generators $L_0$ and $\bar{L}_0$ as
\begin{align}\label{eq:MLrelationship01}
M=\frac{1}{\ell}(L_0+\bar{L}_0),\quad J=L_0-\bar{L}_0.
\end{align}
Therefore, the Cardy formula for the asymptotic growth of states of the dual CFT$_2$ is 
\begin{align}\label{eq:growthstate01}
S_\text{CFT}=2\pi \sqrt{\frac{c n_\text{R}}{6}}+2\pi \sqrt{\frac{c n_\text{L}}{6}},
\end{align}
where $n_\text{R}$ and $n_\text{L}$ are the eigenvalues of $L_0$ and $\bar{L}_0$, respectively. According to Eq.\,(\ref{eq:MLrelationship01}), the Cardy formula in Eq.\,(\ref{eq:growthstate01}) can be rewritten in terms of $M$ and $J$ of the black hole, and it corresponds exactly to the Bekenstein-Hawking entropy \cite{Strominger:1997eq}.
\begin{align}\label{eq:dualentropy03a}
S_\text{CFT}=\pi \sqrt{\frac{\ell(\ell M +J)}{2G}}+\pi \sqrt{\frac{\ell(M\ell-J)}{2G}}=S_\text{BH}.
\end{align}
Therefore, the Bekenstein--Hawking entropy microscopically originates from the asymptotic growth of the states of the CFT$_2$. Considering this correspondence, we herein study the changes in the rotating AdS$_3$ black hole owing to the scattering of a massive scalar field and the impact of these changes in terms of the CFT$_2$.

\section{Solution to Massive Scalar Field}\label{sec3}

We consider that the fluxes of a massive scalar field are scattered by the rotating AdS$_3$ black hole. Owing to the energy and angular momentum of the scalar field that are transferred to the black hole during the scattering, the black hole changes. This energy and angular momentum transferred to the black hole can be estimated based on the fluxes of the scalar field at the horizon. Therefore, according to Eq.\,(\ref{eq:MLrelationship01}), the changes in the black hole can be related to those in the dual CFT$_2$. We follow the general procedure and convention reported in \cite{Gwak:2012hq,Gwak:2018akg,Gwak:2019rcz}. Note that such a scalar solution has also been found in \cite{Chen:2018yah} for the metric of the BTZ black hole\cite{Banados:1992wn,Banados:1992gq}. The action of the massive scalar field begins as
\begin{align}
S_\Psi =-\frac{1}{2}\int d^3 x \sqrt{-g}\left(\partial_\mu \Psi \partial^\mu \Psi^*+\mu^2 \Psi\Psi^*\right),
\end{align}
where we consider the mass of the scalar field $\mu$ to include the null and timelike cases, in terms of a particle. Consequently, the equations of motion for the scalar field are obtained as
\begin{align}
\frac{1}{\sqrt{-g}} \partial_\mu \left(\sqrt{-g} g^{\mu\nu} \partial _\nu \Psi\right)-\mu^2 \Psi=0.
\end{align}
This is rewritten as
\begin{align}
-\frac{1}{N^2}\partial_t^2 \Psi+\frac{2N^\phi}{N^2}\partial_t\partial_\phi\Psi+\frac{1}{r}\partial_r\left(\frac{N^2 \rho^2}{r}\partial_r \Psi\right)+\left(-\frac{(N^\phi)^2}{N^2}+\frac{1}{\rho^2}\right)\partial_\phi^2 \Psi-\mu^2\Psi=0,
\end{align}
where the time and angular parts are easily separated. Thus, we consider the solution of the scalar field with the separation constant $\omega$ and $m$ corresponding to its frequency and angular number.
\begin{align}
\Psi(T,r,\theta,\Phi)=e^{-i\omega t}e^{im\phi} R(r),
\end{align}
The remaining radial equation becomes
\begin{align}\label{eq:radialeq1}
\frac{1}{R(r)}\partial_r\left(\frac{N^2 \rho^2}{r}\partial_r R(r)\right)+\left(\frac{\omega^2}{N^2}+\frac{2\omega m N^\phi}{N^2}+\left(\frac{(N^\phi)^2}{N^2}-\frac{1}{\rho^2}\right)m^2-\mu^2\right)r=0.
\end{align}
The radial equation in Eq.\,(\ref{eq:radialeq1}) can be rewritten and solved in the tortoise coordinate defined as
\begin{align}
\frac{dr^*}{dr}=\frac{r}{N^2 \rho^2}.
\end{align}
The interval $(r_\text{h},+\infty)$ in the $r$ coordinate becomes $(-\infty,0)$ in the tortoise coordinate. The radial equation in Eq.\,(\ref{eq:radialeq1}) is simplified as
\begin{align}
\frac{1}{R(r)}\frac{d^2R(r)}{d{r^*}^2}+\left(\omega^2+2\omega m N^\phi+\left((N^\phi)^2-\frac{N^2}{\rho^2}\right)m^2-\mu^2N^2\right)\rho^2=0.
\end{align}
The energy and angular momentum transferred to the black hole are estimated based on the fluxes of the solution to the scalar field at the horizon. Thus, the solution to the scalar field should be determined at the horizon to obtain these fluxes. The radial equation at the horizon is expressed as
\begin{align}
\frac{1}{R(r)}\frac{d^2R(r)}{d{r^*}^2}+\left(\omega-m \Omega_\text{h}\right)^2\rho_\text{h}^2=0,
\end{align}
where the radial solutions at the horizon are
\begin{align}
R(r)=e^{\pm i(\omega - m \Omega_\text{h})\rho_\text{h} r^*}.
\end{align} 
Here, we assume that the ingoing scalar field is scattered by the black hole; therefore, the radial solution should describe an ingoing wave. The solutions to the ingoing scalar field are
\begin{align}
\Psi=e^{-i\omega t} e^{im \phi} e^{- i(\omega - m \Omega_\text{h})\rho_\text{h}r^*} \text{ : ingoing}, \quad \Psi^*=e^{i\omega t} e^{-im \phi} e^{i (\omega - m \Omega_\text{h})\rho_\text{h}r^*} \text{ : conjugate}.
\end{align}
By applying these solutions, we can obtain the fluxes of energy and angular momentum flowing into the black hole.

\section{Fluxes in Black Hole and States in Dual CFT}\label{sec4}

Based on the fluxes of the scalar field, we can estimate the amount of mass and angular momentum flowing into the rotating AdS$_3$ black hole. Therefore, the changes in the mass and angular momentum of the black hole can be specified for an initial condition. Furthermore, as the conserved quantities of the black hole are directly related to the states $n_\text{R}$ and $n_\text{L}$ in Eq.\,(\ref{eq:MLrelationship01}), these changes are also associated with the variations in the dual CFT$_2$\cite{Gwak:2012hq}.

The fluxes of the scalar field are defined in terms of its energy-momentum tensor, which is expressed as
\begin{align}
T_{\mu\nu}&=\partial_{(\mu}\Psi \partial_{\nu)}\Psi^*-g_{\mu\nu}\left(\frac{1}{2}\partial_\mu\Psi \partial^\mu\Psi^* -\frac{1}{2}\mu^2\Psi\Psi^*\right).
\end{align}
The energy and angular momentum fluxes are obtained as 
\begin{align}\label{eq:fluxes01}
\frac{dE}{dt}=\int T_T^r \sqrt{-g} d\Phi=2\pi\omega(\omega-m\Omega_\text{h})\rho_\text{h},\,\,\frac{dJ}{dt}=-\int T^r_\phi \sqrt{-g}d\Phi=2\pi m(\omega-m\Omega_\text{h})\rho_\text{h},\,\,\frac{dJ}{dt}&=\frac{\omega}{m} \frac{dE}{dt}.
\end{align}
The energy and angular momentum of the scalar field are conserved quantities. As they are adding to the corresponding quantities of the black hole, mass and angular momentum, they are conserved. According to the fluxes in Eq.\,(\ref{eq:fluxes01}), we can elucidate the exact changes in the mass and angular momentum during the infinitesimal time interval $dt$.  
\begin{align}\label{eq:MLrelationship03}
dM=2\pi\omega(\omega-m\Omega_\text{h})\rho_\text{h}dt,\quad dJ=2\pi m(\omega-m\Omega_\text{h})\rho_\text{h}dt.
\end{align}
Thus, we can determine the variation of the eigenvalues in the dual CFT, under these changes. The eigenvalues are related to the mass and angular momentum of the black hole, as shown in Eq.\,(\ref{eq:MLrelationship01}). Hence, we obtain the relationship between the changes in the black hole and the dual CFT.
\begin{align}\label{eq:MLrelationship02}
d n_\text{R}=\pi (\omega -m\Omega_\text{h})(\omega \ell +m)\rho_\text{h}dt,\quad d n_\text{L}=\pi (\omega -m\Omega_\text{h})(\omega \ell -m)\rho_\text{h}dt.
\end{align}
Eq.\,(\ref{eq:MLrelationship02}) provides the tendency of the variations in the dual CFT$_2$ originating from the changes in the black hole caused by the scattering of the scalar field.

\section{Variations in Dual CFT$_2$}\label{sec5}

We discuss the bounded behaviors of the rotating AdS$_3$ black hole and their implications for the dual CFT$_2$. Owing to the transferred energy and angular momentum, the mass and angular momentum of the black hole changes under constraints such as the laws of thermodynamics. Hence, the changes in the black hole are related to the variations in the eigenvalues for the dual CFT$_2$, which can now be analyzed.

\subsection{Non-Zero States}

\begin{figure}[h]
\centering
\subfigure[{For $\omega= 1$, $m=0$.}] {\includegraphics[scale=0.51,keepaspectratio]{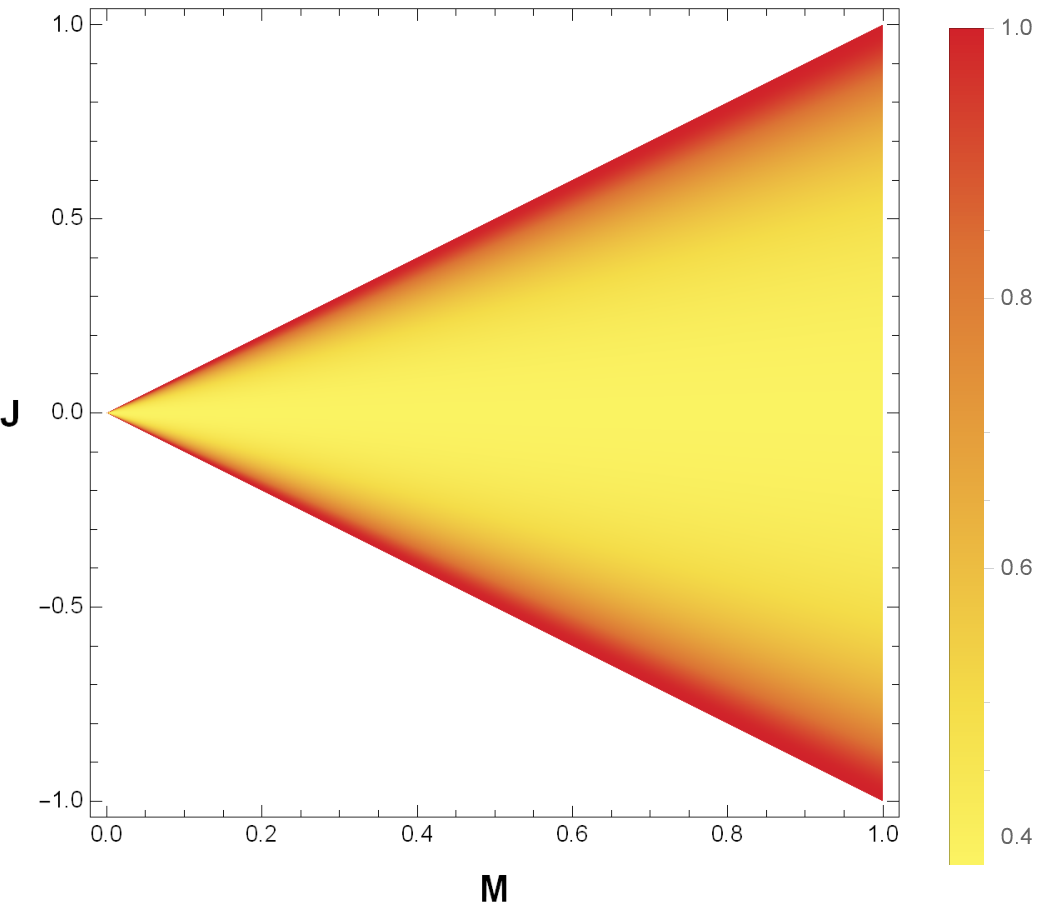}}\quad
\subfigure[{For $\omega= 1$, $m= 2$.}] {\includegraphics[scale=0.51,keepaspectratio]{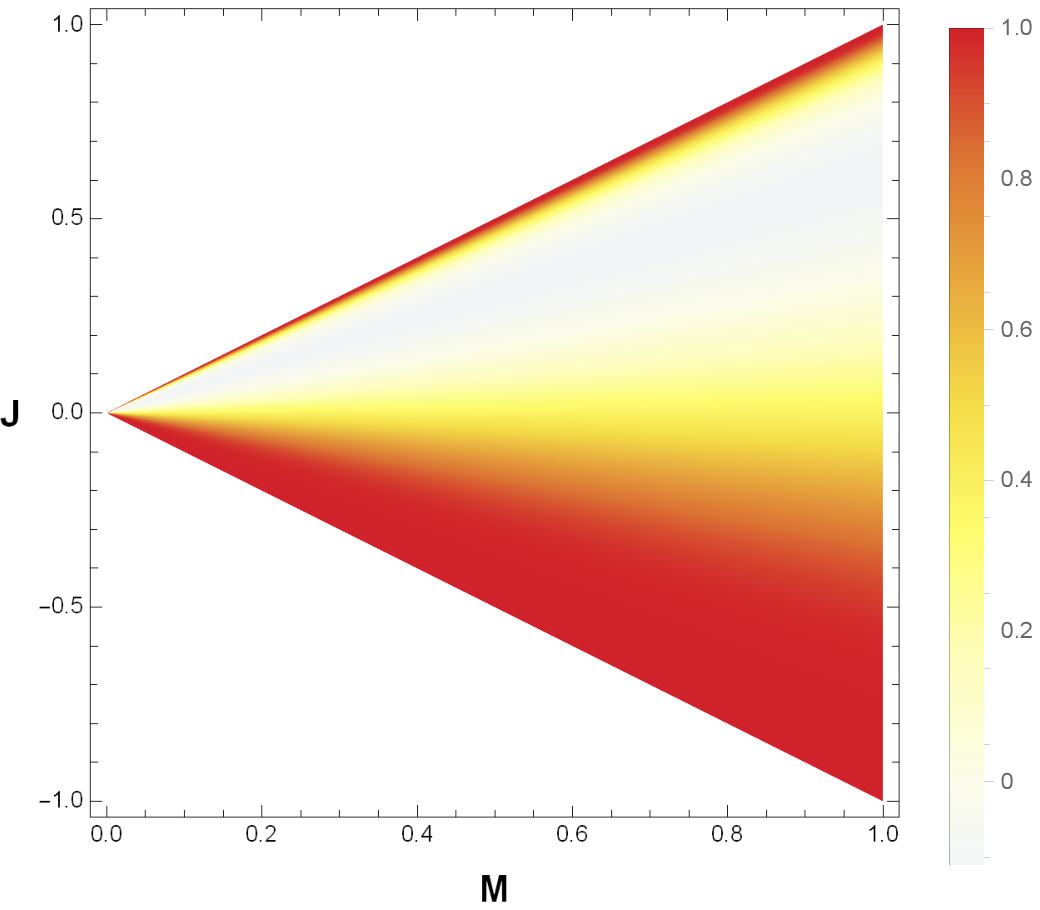}}\quad
\subfigure[{For $\omega= 2$, $m= 3$.}] {\includegraphics[scale=0.51,keepaspectratio]{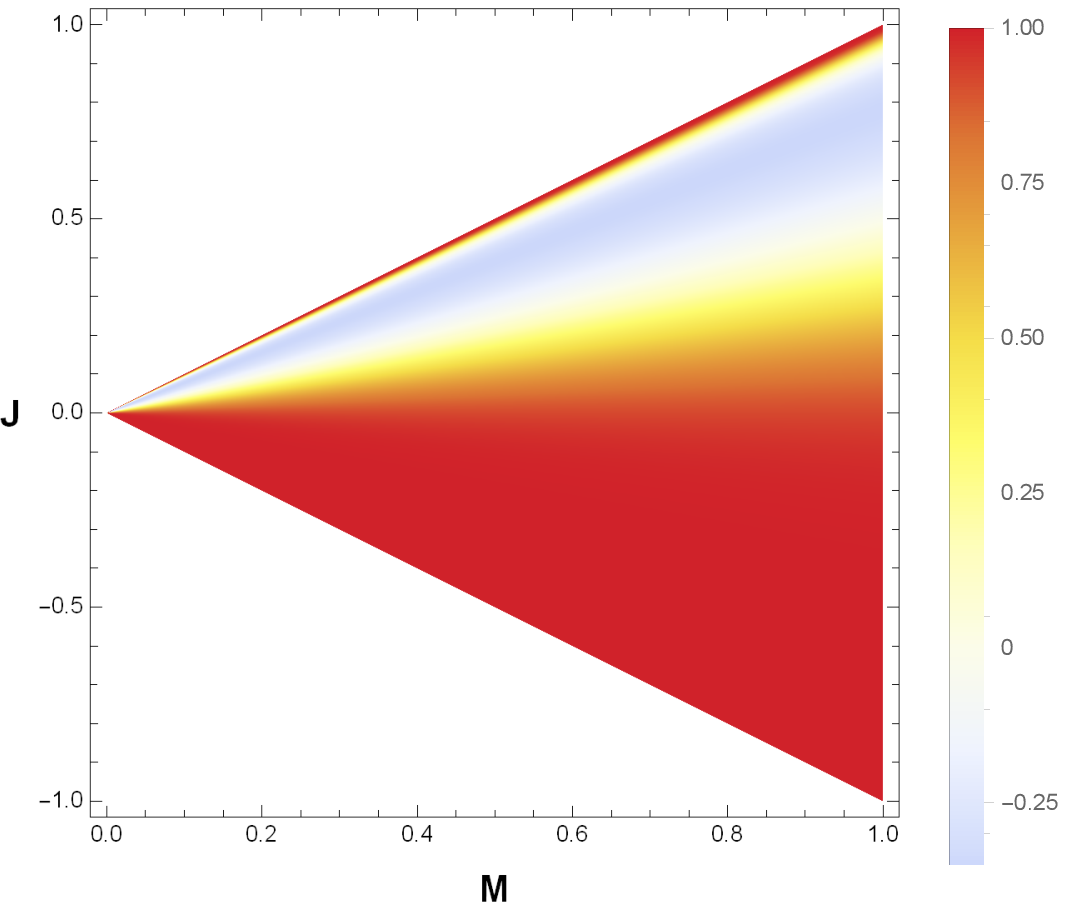}}
\caption{{\small Graphs for $\tanh\left(\frac{1}{300}\frac{dT_\text{H}}{dt}\right)$ in rotating AdS$_3$ black holes with the mode $(\omega,m)$ of the scalar field.}}
\label{fig:fig1}
\end{figure}

The mass and angular momentum transferred by the fluxes of the scalar field alter various properties of the black hole. Among these properties, the Hawking temperature is a relevant quantity for measuring the tendency of changes depending on the scalar field. Owing to these fluxes, the temperature $T_\text{H}(M,J)$ changes during the infinitesimal time interval.
\begin{align}
T_\text{H}(M+dM,J+dJ,r_\text{h})=\frac{\partial T_\text{H}}{\partial M}dM+\frac{\partial T_\text{H}}{\partial J}dJ,
\end{align}
where
\begin{align}
\frac{\partial T_\text{H}}{\partial M}=\frac{4GM\ell}{\pi \rho_\text{h} \sqrt{M^2\ell^2-J^2}}-\frac{2G}{\pi \rho_\text{h}},\quad \frac{\partial T_\text{H}}{\partial J}=\frac{8GJ}{\pi \rho^3}-\frac{4GJ}{\pi \ell \rho_\text{h}\sqrt{M^2\ell^2 -J^2}}.\nonumber
\end{align}
Using Eq.\,(\ref{eq:MLrelationship03}), we can determine the change in the temperature during the infinitesimal time interval.
\begin{align}\label{eq:temp01}
dT_\text{H}=\frac{4G\ell (4GJ^2+M\rho_\text{h}^2)}{\rho_\text{h}^2 \sqrt{M^2\ell^2-J^2}}(\omega-m\Omega_\text{h})\left(\omega-m\frac{J(4GM\ell^2+\rho_\text{h}^2)}{\ell^2(4GJ^2+M\rho_\text{h}^2)}\right)dt.
\end{align}
The change in Eq.\,(\ref{eq:temp01}) depends on the initial state of the black hole as well as the modes of the scalar field. Therefore, the temperature may increase or decrease. However, when the initial black hole becomes extremal, the change in the temperature has a tendency. To denote the extremality of the black hole, we introduce
\begin{align}
\delta\equiv M^2\ell^2-J^2.
\end{align}
When $\delta$ is zero, the black hole is extremal. As the change in Eq.\,(\ref{eq:temp01}) is divergent to the extremal black hole, it can be expanded under the near-extremal condition, $\delta\ll 1$\cite{Gwak:2019rcz}. Thus, Eq.\,(\ref{eq:temp01}) becomes 
\begin{align}\label{eq:temp02}
dT_\text{H}=\frac{8GM\ell}{\sqrt{\delta}}(\omega-m\Omega_\text{e})^2dt+\mathcal{O}(\delta^0),\quad \Omega_\text{e}=\frac{J}{M\ell^2},
\end{align}
where $\Omega_\text{e}$ is the angular velocity for the extremal black hole. When the black hole becomes extremal, the leading term in Eq.\,(\ref{eq:temp02}) becomes dominant and positive for all modes of the scalar field. This implies that the temperature of the near-extremal black hole always increases. However, the temperature of an extremal black hole is zero. Therefore, a non-extremal black hole cannot evolve to an extremal black hole under this process. This is in agreement with the third law of thermodynamics; in other words, a black hole cannot convert to the extremal state via a physical process. Thus, we conclude that the third law of thermodynamics is still valid, using the scattering of the scalar field. Details regarding the temperature are depicted in Fig.\,\ref{fig:fig1}. 
Here, each point $(M,J)$ represents the initial state of the black holes, and the colors represent the temperature changes for the given modes $(\omega,m)$ of the scalar field. Compared to the value of the change, the sign of the change is more important; therefore, we preserve the sign of temperature change using a hyperbolic tangent function. It should be noted that the denominator $300$ is arbitrarily chosen.  According to Fig.\,\ref{fig:fig1}, changes in the temperature can be positive or negative depending on the initial states. In particular, when the initial states are close to the extremal condition, sudden and large positive changes occur, as presented in Eq.\,(\ref{eq:temp02}). Thus, we reveal that a black hole cannot be extremal, based on the third law of thermodynamics. This implies that the extremal condition works as a type of bound to attaining the extremal condition, in the parameter space $(M,J)$. We show the significance of the bounds in the dual CFT$_2$, according to the AdS/CFT correspondence. 

\begin{figure}[h]
\centering
\subfigure[{For $\omega= 1$, $m=0$.}] {\includegraphics[scale=0.51,keepaspectratio]{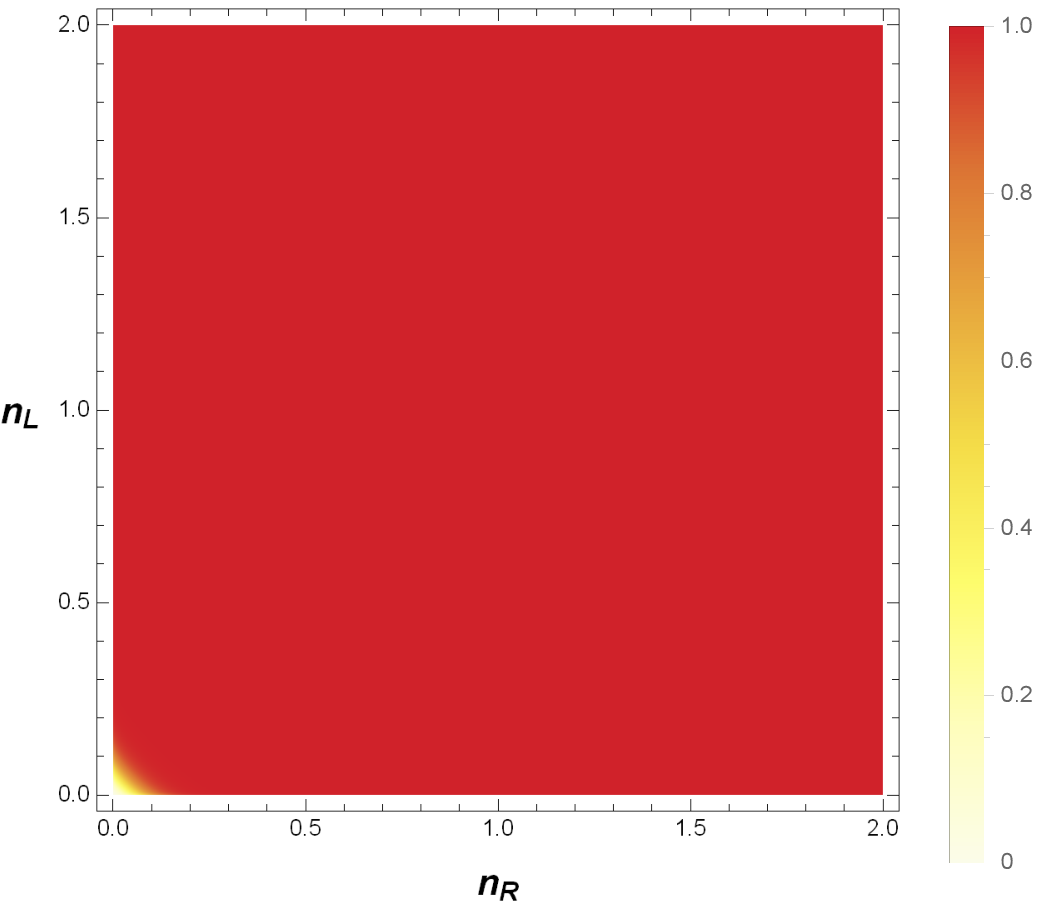}}\quad
\subfigure[{For $\omega= 1$, $m= 2$.}] {\includegraphics[scale=0.51,keepaspectratio]{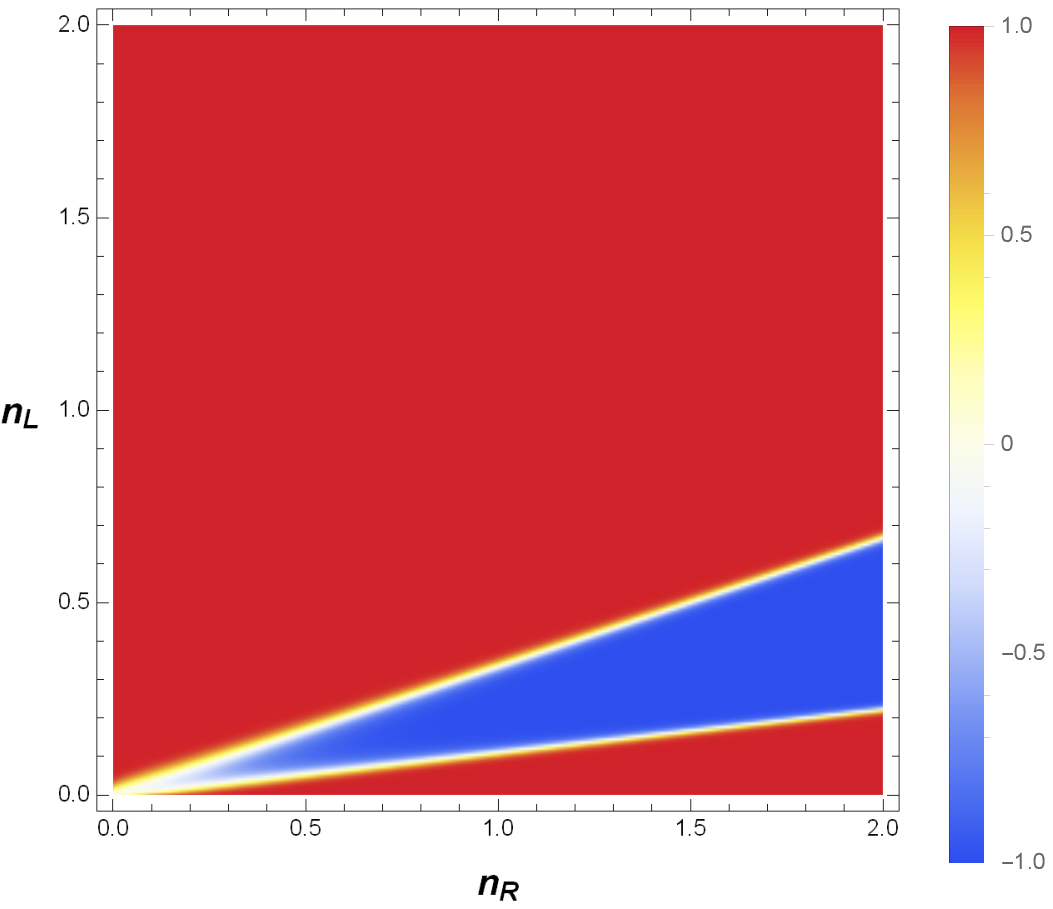}}\quad
\subfigure[{For $\omega= 2$, $m= 3$.}] {\includegraphics[scale=0.51,keepaspectratio]{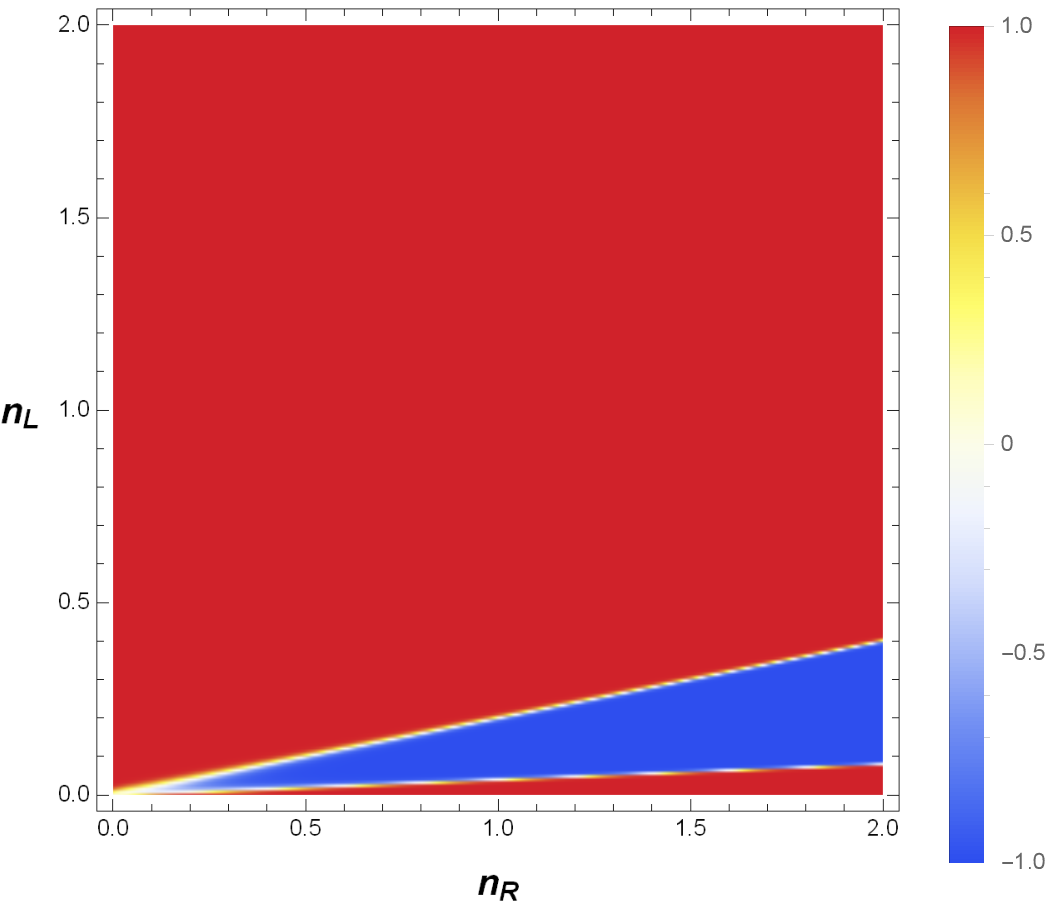}}
\caption{{\small Graphs for $\tanh\left(\frac{d\delta}{dt}\right)$ in eigenvalues $(n_\text{R},n_\text{L})$ with the mode $(\omega,m)$ of the scalar field.}}
\label{fig:fig2}
\end{figure}

The third law for the black hole side acts as a bound for the eigenvalues $(n_\text{L},n_\text{R})$. According to the third law, the extremal condition cannot be saturated via a physical process. In terms of the eigenvalues related to the dual CFT, the extremal condition is rewritten as
\begin{align}
\delta = (n_\text{R}+n_\text{L})^2-(n_\text{R}-n_\text{L})^2.
\end{align}
To make $\delta$ equal to zero, $n_\text{R}$ and/or $n_\text{R}$ should be zero. As the third law prevents $\delta$ from being equal to zero, neither $n_\text{R}$ nor $n_\text{R}$ can be zero for a process. For the eigenvalues corresponding to the non-extremal black holes, the variation of $\delta$ is
\begin{align}
d\delta&=\frac{\partial \delta}{\partial n_\text{R}}dn_\text{R}+\frac{\partial \delta}{\partial n_\text{L}}dn_\text{L}\\
&=\frac{8\pi (m(n_\text{L}-n_\text{R})+(n_\text{L}+n_\text{R})\ell \omega)\left(m(n_\text{L}-n_\text{R})+\left(n_\text{L}+n_\text{R}+2\sqrt{n_\text{L}n_\text{R}}\right)\ell\omega\right)}{\sqrt{\left(n_\text{L}+n_\text{L}+2\sqrt{n_\text{L} n_\text{R}}\right)\ell}}.\nonumber
\end{align}
This can be seen in Fig.\,\ref{fig:fig2}, which is rescaled using the hyperbolic tangent to preserve the sign of the change in temperature.
As the change in $\delta$ is positive near $n_\text{R}=0$ or $n_\text{L}=0$, the eigenvalues cannot be zero under the variations. Interestingly, this implies that the black hole cannot be $M=0$ vacuum by extracting its energy, because $n_\text{R}=n_\text{R}=0$ cannot be achieved, as previously stated. The black hole is considered to be excited from $M=0$ vacuum\cite{Strominger:1997eq}; hence, $n_\text{R}=n_\text{R}=0$ also acts as a bound. Therefore, the eigenstates of the dual CFT cannot become zero owing to any process.

\subsection{Asymptotic Growth of States}

The Cardy formula indicates that the asymptotic growth of the states corresponds to the Bekenstein--Hawking entropy, as shown in Eq.\,(\ref{eq:dualentropy03a}). When the mass and angular momentum of the rotating AdS$_3$ black hole is changed owing to the scalar field, the dual CFT$_2$ alters the eigenvalues $n_\text{R}$ and $n_\text{L}$. Consequently, this changes the value of the asymptotic growth of states. According to the variation in the asymptotic growth of states, we determine the tendency of the eigenvalues and the application of the second law in the dual CFT. The asymptotic growth of states varies as 
\begin{align}
dS_\text{CFT}=\frac{\partial S_\text{CFT}}{\partial n_\text{R}}dn_\text{R}+\frac{\partial S_\text{CFT}}{\partial n_\text{L}}dn_\text{L},
\end{align}
where the variations in the eigenvalues are given in Eq.\,(\ref{eq:MLrelationship02}) in terms of the parameters of the scalar field. Thus, we obtain 
\begin{align}\label{eq:asymptoticgrowth05}
dS_\text{CFT}=\pi^2\ell \left(\sqrt{\frac{c}{6n_\text{R}}}+\sqrt{\frac{c}{6n_\text{L}}}\right)(\omega-m\Omega_\text{h})^2 \rho_\text{h}dt\geq 0.
\end{align}
The increase in the asymptotic growth of states represents the second law of thermodynamics, according to Eq.\,(\ref{eq:dualentropy03a}). Furthermore, because we already know that $n_\text{R}$ and $n_\text{L}$ are non-zero values, the variation in the Cardy formula will diversify near $n_\text{R}=0$ or $n_\text{L}=0$.
\begin{figure}[h]
\centering
\subfigure[{For $\omega= 1$, $m=0$.}] {\includegraphics[scale=0.51,keepaspectratio]{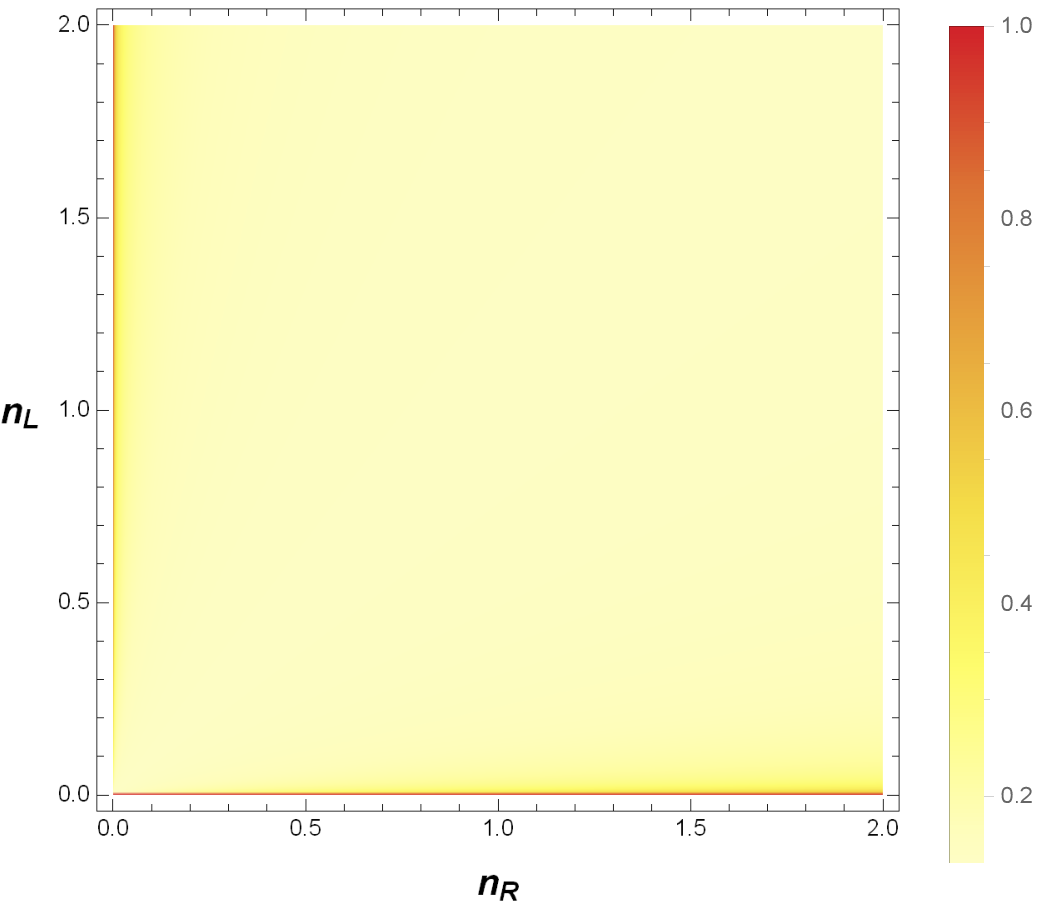}}\quad
\subfigure[{For $\omega= 1$, $m= 2$.}] {\includegraphics[scale=0.51,keepaspectratio]{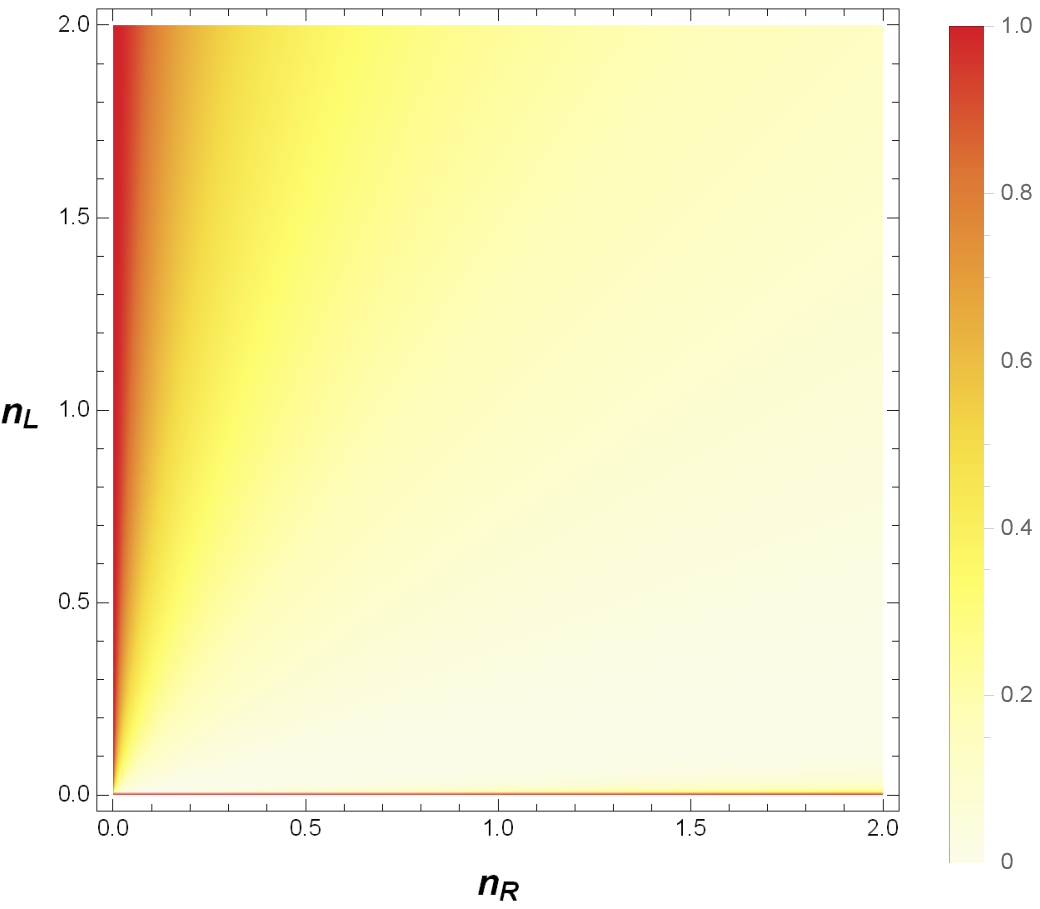}}\quad
\subfigure[{For $\omega= 2$, $m= 3$.}] {\includegraphics[scale=0.51,keepaspectratio]{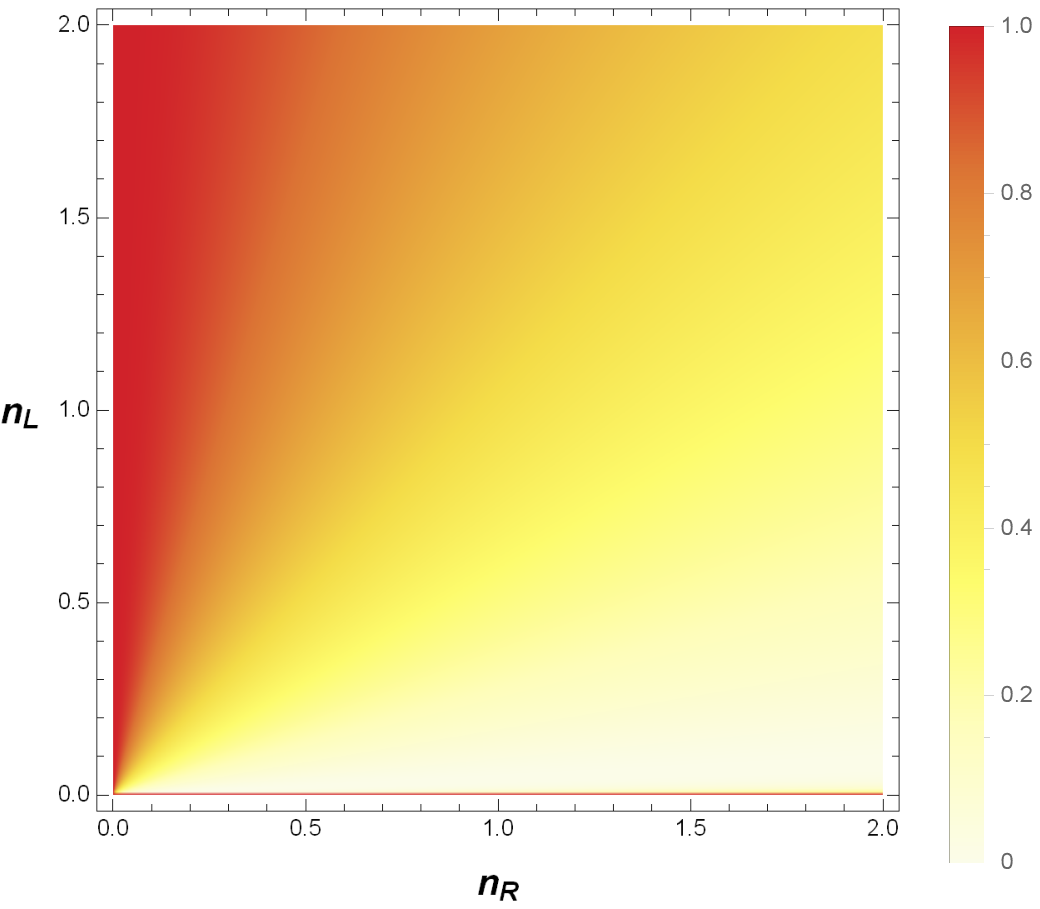}}
\caption{{\small Graphs for $\tanh\left(\frac{1}{300}\frac{dS_\text{CFT}}{dt}\right)$ in eigenvalues $(n_\text{R},n_\text{L})$ with the mode $(\omega,m)$ of the scalar field.}}
\label{fig:fig3}
\end{figure}
The detailed behaviors in Eq.\,(\ref{eq:asymptoticgrowth05}) are depicted given in Fig.\,\ref{fig:fig3}. Here, we ensure an increase in the asymptotic growth. Furthermore, as shown in Fig.\,\ref{fig:fig2}, the eigenvalues cannot be zero, and the asymptotic growth increases sharply, similar to the bound, to ensure non-zero eigenvalues.

Therefore, according to the laws of thermodynamics for the black hole, the eigenvalues $n_\text{R}$ and $n_\text{L}$ cannot be zero. If one of these eigenvalues tends to zero, the asymptotic growth increases sharply, thereby increasing the eigenvalue.

\section{Summary}\label{sec6}

We studied the relationships between the physical boundaries for the rotating AdS$_3$ black hole and its dual CFT$_2$, in terms of the AdS/CFT correspondence. The black hole is specified by its conserved quantities. The conserved quantities, such as mass and angular momentum, are clearly bounded by the extremal condition. The third law of thermodynamics ensures that extremality cannot be achieved via any physical process, and the second law implies that such a bounded behavior is naturally preferred. For the rotating AdS$_3$ black hole satisfying this particular boundary condition, conserved quantities can be directly related to the microstates of its dual CFT$_2$. Moreover, the second law of thermodynamics coincides with the Cardy formula in the dual CFT$_2$. Owing to the scattering of the scalar field, we conclude that changes in the black hole conform to the laws of thermodynamics and that the third and second laws act as bounds to the mass and angular momentum. Consequently, we imposed this particular relationship for a few black holes to rewrite the conserved quantities as eigenvalues of the dual CFT$_2$. Here, the third law implies that the eigenvalues $n_\text{R}$ and $n_\text{L}$ cannot be zero. The variation in the eigenvalues decreases their difference; hence, the extremality becomes small under the changes caused by the scalar field. The second law ensures that the asymptotic growth of states in the CFT$_2$ side increases sharply when one of the eigenvalues tends to zero. Therefore, such an extremely large difference between $n_\text{R}$ and $n_\text{L}$ cannot be achieved via a physical process. Even if such a large difference is considered as an initial condition, it decreases by a small perturbation. Therefore, according to the AdS/CFT correspondence, the laws of thermodynamics for the rotating AdS$_3$ black hole are closely related to the non-zero state of the eigenvalues in the CFT$_2$.

\vspace{10pt}

\noindent{\bf Acknowledgments}

\noindent This work was supported by the National Research Foundation of Korea (NRF) grant funded by the Korea government (MSIT) (NRF-2018R1C1B6004349) and the Dongguk University Research Fund of 2020.

\end{document}